%% file: btap10-jsbm.tex
\documentclass{llncs}

\usepackage{amsmath}		
\usepackage{amsfonts}		
\usepackage{amssymb}		

\usepackage{graphicx}
\usepackage{pstricks}
\usepackage{pst-node}
\usepackage{multirow}
\usepackage{color}

\usepackage{epsfig}

\usepackage{xspace}

\usepackage{bsymb}

\usepackage{multicol}
\input{macros.tex}

\title{Syntactic Abstraction of B Models\\to Generate Tests} 

\author{
J. Julliand$^1$ \and N. Stouls$^2$ \and P.-C. Bu\'e$^1$ \and P.-A. Masson$^1$}

\institute{$^1$~LIFC, Universit\'e de Franche-Comt\'e\\
16, route de Gray F-25030 Besan\c{c}on Cedex\\
\texttt{\{bue,\hspace{0.5ex}julliand,\hspace{0.5ex}masson\}@lifc.univ-fcomte.fr}\\[3pt]
$^2$~Universit\'e de Lyon, INRIA\\
INSA-Lyon, CITI, F-69621, France\\
\texttt{nicolas.stouls@insa-lyon.fr}
}

\begin{document}

\maketitle

\begin{abstract}
%
%
%
  In a model-based testing approach as well as for the verification of
  properties, B models provide an interesting
  solution. However, for industrial applications, the size of their state
  space often makes them hard to handle. To reduce the amount of
  states, an abstraction function can be used, often combining state
  variable elimination and domain abstractions of the remaining
  variables. This paper complements previous results, based on domain
  abstraction for test generation, by adding a preliminary syntactic  abstraction phase,
  based on variable elimination.  
 We define a syntactic transformation that 
  suppresses some variables from a \B event model, in 
  addition to a method that chooses relevant variables 
  according to a test purpose. We propose two methods to compute 
an abstraction A of an initial model M. The first one computes A as a simulation
of M, and the second one computes A as a bisimulation of M.
The abstraction process produces a finite state system.
We apply this abstraction computation to a Model Based Testing
process.

\mbox{}

\end{abstract}


\input{sec1-introduction.tex}
\input{sec2-preliminaries.tex}

\input{sec3-AlimSysCaseStudy.tex}

\input{sec4-SyntacticAbstraction.tex}

\input{sec5-AbstractionProcess.tex}

\input{sec6-experimentalResults.tex}

\input{sec7-conclusion.tex}

\bibliographystyle{splncs}
\bibliography{secTestBiblio}

\ifLNCSVersion{}{
\clearpage
\appendix
\input{Annexe-ModX}
\input{Annexe-proof_th1}

\input{Annexe-proof2}
}
\end{document}

%% file: macros.tex
\newcommand{\fm}{\textsf{M}\xspace} 

\newcommand{\Pred}{\pred} 

\newcommand{\bdef}{~\hat =~}    

\newcommand{\genesyst}{\emph{GeneSyst}\xspace}

\newcommand{\abs}{\textsf{A}\xspace}
\newcommand{\raf}{\textsf{R}\xspace}

\newcommand{\tp}{\textsf{TP}\xspace}

\def \B       {B\xspace}
\def \CNF     {CF\xspace}
\def \PO		  {PO\xspace}
\def \BES	    {\langle X, I, Init, Ev \rangle}
\def \BESM	  {\langle 
	X_\fm, I_\fm, Init_\fm, Ev_\fm \rangle}
\def \BESA		{\langle 
	X_\abs, I_\abs, Init_\abs, Ev_\abs \rangle}

\def \A				{\mathcal{A}}
\def \AA      {\A_\abs}
\def \AM      {\A_\fm}
\def \Pred    {\mathcal{P}red}

\def \evname {ev}
\def \Ev      {\textit{Ev}}

\newcommand{\ifLNCSVersion}[2]{#2} 

%% file: sec1-introduction.tex
\section{Introduction}
\label{sec:intro}

\B models are well suited for producing tests of an implementation by
means of a \emph{model-based testing}
approach~\cite{MBTofReacSys05,MBTbook06} and to verify dynamic
properties by model-checking~\cite{ProB}. But model-checking as well
as test generation require the models to be finite, and of tractable
size. This is not usually the case with industrial applications, for
which the exploration of the executions modelled frequently comes up
against combinatorial explosion problems.
Abstraction techniques allow for projecting the (possibly infinite or
very large) state space of a system onto a small finite set of
symbolic states. Abstract models make test generation or
model-checking possible in practice~\cite{bcdg-B2007}.
In~\cite{bbjm-amost10}, we have proposed and experimented with an approach of test
generation from abstract models. It appeared that the computation time
of the abstraction could be very expensive, as evidenced by the
Demoney~\cite{specDemoney-tl-02}
case study.
We had replaced a problem of time for searching in a state graph with
a problem of time for solving proofs, as the abstraction was computed
by proving enabledness and reachability conditions on symbolic
states~\cite{genesystZB05}.

In this paper, we contribute to solving this proving time problem by
defining a syntactic abstraction function that requires no proof.
Inspired from slicing techniques~\cite{Weiser84},
%
the function works by suppressing some state
variables from a model.
In order to produce a state system that is both finite and
sufficiently small, we still have to perform a semantic
abstraction. This requires that some proof obligations are solved, but
there are less of them than with the initial model, since it has been
syntactically simplified.
%
This approach results in semantic pruning of generated proof obligations
as proposed in~\cite{CGS09}.

In Sec.~\ref{sec:preliminaries}, we introduce the notion of \B event
system and some of the main properties
of substitution computation. 
Section~\ref{sec:ElecSystem} presents an Electrical System case study
that illustrates our approach. In Sec.~\ref{sec:SyntacticAbstraction},
we first define the set of variables to be preserved by the
abstraction function and then we define the abstraction function
itself.
We prove that this function is correct in the sense that the generated
abstract model \abs simulates or bisimulates the initial model \fm.
In this way, the abstraction can be used to verify safety properties
and to generate tests. In Sec.~\ref{sec:process}, we present an end to
end process to compute test cases from a set of observed variables by
using both the semantic and the syntactic abstractions.
In Sec.~\ref{sec:expResults}, we compare this process to a completely
semantic one on several examples, and we evaluate the practical
interest for test cases generation.  Section~\ref{sec:conclusion}
concludes the paper, gives some future research directions and
compares our approach to other abstraction methods.

%% file: sec2-preliminaries.tex
\section{\B Event Systems and Refinement}
\label{sec:preliminaries}
We use the \B notation~\cite{Bbook} to describe our models: this
section gives the background required for reading the paper. 
Let us first define the following \B notions: primitive forms of
substitution, substitution properties and refinement.  Then we will
summarize the principles of before-after predicates, and conjunctive
form (\CNF) of \B predicates.

First introduced by J.-R. {\sc Abrial}~\cite{abrial-eventB-96}, a \B
event system defines a closed specification of a system by a set of
events.
In the sequel, we use the following notations: $x$, $x_i$,
$y$, $z$ are 
variables and $X$, $Y$, $Z$ are sets of variables.
$\Pred$ is the set of \B 
predicates. 
$I$ ($\in \Pred$) is an invariant, and $P$, $P_1$ and $P_2$ ($\in
\Pred$) denote other predicates. The modifications of the variables
are called \emph{substitutions} in \B, following~\cite{Hoare69} where
the semantics of an assignment is defined as a substitution. In \B,
substitutions are \emph{generalized}: they are 
the semantics of every kind of action, as expressed
by formulas~\ref{eqn1} to~\ref{eqn4} below.
We use $S$, $S_1$ and $S_2$ to denote \B generalized substitutions,
and $E$, $E_i$ and $F$ to denote \B expressions.
%
The \B events are defined as generalized substitutions.  
%
All the substitutions allowed in \B event systems can be rewritten by
means of the five \B primitive forms of substitutions of
Def.~\ref{DefPrimitiveSubst}. Notice that the multiple assignment
can be generalized to $n$ variables.
It is commutative, i.e. $x,~y~:=~E,~F \defi y,~x~:=~F,~E$.

 
%

\begin{definition}[Substitution] 
\label{DefPrimitiveSubst}
The following five substitutions are primitive:
\begin{itemize}
\item single and multiple assignments, denoted as $x~:=~E$ and $x,~y~:=~E,~F$ 

\item substitution with no effect, denoted as $skip$

\item guarded substitution,
  denoted as $P \Rightarrow S$

\item bounded nondeterministic choice,
  denoted as $S_1 [] S_2$

\item substitution with local variable $z$,
  denoted as $@z.S$.
\end{itemize}
\end{definition}

Notice that the substitution with local variable is mainly used to express
the unbounded nondeterministic choice 
denoted by $@z.(P \Rightarrow S)$. 
Let us specify that 
among the usual structures of specification
languages, the conditional substitution \textsc{IF} $P$ \textsc{THEN}
$S_1$ \textsc{ELSE} $S_2$ \textsc{END} is denoted by $(P \Rightarrow
S_1) [] (\neg P \Rightarrow S_2)$ with the primitive forms.


Given a substitution $S$ and a post-condition $P$, it is possible to
compute the weakest precondition such that if it is satisfied, then
$P$ is satisfied after the execution of $S$.  The weakest
precondition is denoted by $[S]P$. $[x~:=~E]P$ is the usual
substitution of all the free occurrences of $x$ in $P$ by $E$. For the four
other primitive forms, the weakest precondition is computed as
indicated by formulas~\ref{eqn1} to~\ref{eqn4} below, proved
in~\cite{Bbook}.


\begin{small}\vspace*{-10pt}
\begin{eqnarray}
\label{eqn1}
&[skip]P \Leftrightarrow P\\
\label{eqn2}
&[P_1 \Rightarrow S]P_2 \Leftrightarrow (P_1 \Rightarrow [S]P_2)\\
\label{eqn3}
&[S_1 [] S_2]P \Leftrightarrow [S_1]P \wedge [S_2]P\\
\label{eqn4}
&[@z.S]P \Leftrightarrow \forall z.[S]P &\textrm{ if } z \textrm{ is
  not  free in } P\\
\label{eqn5}
\textrm{Distributivity: }& [S](P_1 \wedge P_2) \Leftrightarrow [S]P_1
\wedge [S]P_2
\end{eqnarray}\vspace*{-10pt}
\end{small}

Definition~\ref{DefBeventSystem} defines correct \B event systems.
To explicitly refer to a given model, we add the name of that model as
a subscript to the symbols $X$, $I$, $Init$ and $\Ev$. $I_\fm$ is for
example the invariant of a model $\fm$.



\begin{definition}[Correct \B Event System] 
\label{DefBeventSystem}
A correct \B event system is a tuple $\BES$ where:
\begin{itemize}


\item $X$ is a set of state variables,

\item $I$ ($\in\Pred$) is an invariant predicate over $X$,

\item $\textrm{Init}$ is a substitution called \emph{initialization}, 
such that the invariant holds in any initial state:
 $[Init]I$, 

\item $Ev$ is a set of event definitions in the shape of $\evname_i
  \defi S_i$ such that every event preserve the invariant:
   $I \Rightarrow [S_i]I$.
\end{itemize}
\end{definition}


In Sec.~\ref{sec:SyntacticAbstraction}, we will prove that an
abstraction \abs that we compute is refined by its source event system
\fm, and so we give in Def.~\ref{DefBeventSystemRefinement} the
definition of a \B event system refinement.  

\begin{definition}[\B Event System Refinement] 
\label{DefBeventSystemRefinement}
Let $\abs$ and $\raf$ be two correct \B event systems.  Let $I_\raf$ be their
gluing invariant,
i.e. a predicate that
indicates how the values of the variables in $\raf$ and $\abs$ relate
to each other.  \raf refines \abs if:
\begin{itemize}
\item any initialization of $\raf$ is associated to an initialization of $\abs$ according to $I_\raf$: 
      ${[Init_\raf] \neg [Init_\abs] \neg I_\raf}$

\item any event $ev \defi S_\raf$ of $\raf$ is an event of $\abs$ defined by $ev \defi S_\abs$ in $\Ev_\abs$ that
satisfy $I_\raf$: $
I_\abs \wedge I_\raf \Rightarrow [S_\raf] \neg [S_\abs] \neg I_\raf$.

\end{itemize}
\end{definition}




This paper also relies on two more definitions:
the before-after predicate and the \CNF form.  We denote by $Prd_X(S)$
the before-after predicate of a substitution $S$. It defines the
relation between the values of the variables of the set $X$ before and
after the substitution $S$.  A primed variable denotes its after
value.
From~\cite{Bbook}, the before-after predicate is defined by:

\begin{small}\vspace*{-10pt}
\begin{eqnarray}
\label{eqn6}
Prd_X(S) \defi \neg [S] \neg (\bigwedge_{x \in X}(x=x')).
\end{eqnarray}\vspace*{-10pt}
\end{small}

 
\begin{definition}[Conjunctive Form] 
\label{DefCNF}
A \B predicate $P \in \Pred$ is in \CNF when it is a conjunction $p_1
\wedge p_2 \wedge \ldots \wedge p_n$ where every $p_i$ is a
disjunction $p_i^1 \vee p_i^2 \vee \ldots \vee p_i^m$ such that any
$p_i^j$ is an elementary predicate in one of the following two forms:
\begin{itemize}
\item 
  $E(Y)~r~F(Z)$, where $E(Y)$ and $F(Z)$ are \B
  expressions on the sets of variables $Y$ and $Z$ and $r$ is a
  relational operator,

\item $\forall z.P$ or $\exists z.P$, where $P$ is a \B predicate in \CNF.
\end{itemize}
\end{definition}


Section~\ref{sec:SyntacticAbstraction} will define predicate
transformation rules.  We put the predicates in \CNF according to
Def.~\ref{DefCNF} before their transformation.
This allows the transformation to be correct although the
negation is not monotonic w.r.t a transformation $T$ of the
predicates: $T(\neg P) \neq \neg T(P)$.

%% file: sec3-AlimSysCaseStudy.tex
\section{Electrical System Example}
\label{sec:ElecSystem}
We describe in this section a \B event system that we will use in this
paper as a running example to illustrate our proposal.

\begin{figure}
	\centering
	\includegraphics[width=5cm]{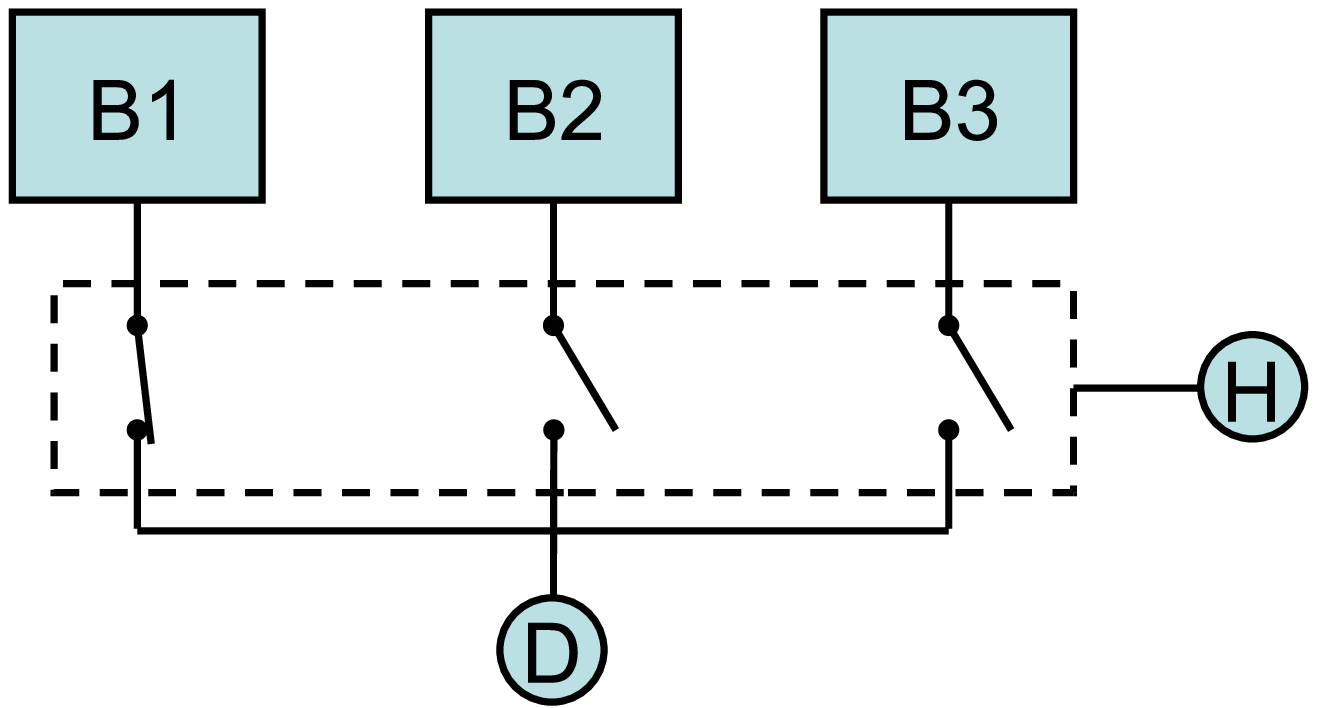}
	\caption{Electrical System\label{fig:ElectricalSystem}}
\end{figure}

A device D is powered by one of three batteries $B_1, B_2, B_3$ as
shown in Fig.~\ref{fig:ElectricalSystem}. A switch connects (or not) a
battery $B_i$ to the device D.  A clock H periodically sends a signal
that causes a commutation of the switches, i.e. a change of the
battery in charge of powering the device D. The working of the system
must satisfy the three following requirements:
\begin{itemize}
\item $Req_1$: 
no short-circuit, i.e. there is only one switch 
  closed at a time,
\item $Req_2$: 
continuous power supply, i.e. there is always
  one switch closed, 
\item $Req_3$: a signal from the clock always changes the switch that
  is closed.
\end{itemize}

The batteries are subject to electrical failures. If it occurs
to the battery that is powering D, the system triggers an exceptional
commutation to satisfy the requirement $Req_2$. The broken batteries
are replaced by a maintenance service. We assume that it works fast
enough for not having more than two batteries down at the same time.
When two batteries are down, the requirement $Req_3$ is relaxed
and the clock signal leaves unchanged the switch that is closed.

This system is modeled in Fig.~\ref{fig:BSpecElecSystem} 
by means of three variables. 
$H$ models the clock and takes two values: $tic$ when it asks
for a commutation and $tac$ when this commutation has occurred. 
$Sw$ models the state of the three switches by an
integer between $1$ and $3$: $Sw=i$ indicates that the switch~$i$ is
closed while the others are opened. This modelling makes that
requirements $Req_1$ and $Req_2$ necessarily hold.
$Bat$ models the electrical failures by a total function. 
The $ko$ value for a battery indicates that
it is down. In addition to the typing of the variables, the invariant
$I$ expresses 
the assumption that at least one
battery is not down by stating that 
$Bat(Sw)=ok$.
Notice that the requirement $Req_3$ is a dynamic property, not
formalized in~$I$. The initial state is defined by $Init$ in
Fig.~\ref{fig:BSpecElecSystem}.  The behavior of the system is
described by four events: 
\begin{itemize}
\item \textbf{Tic} sends a commutation command,

\item \textbf{Com\footnote{An expression $r \ranres E$ denotes a relation 
where the range is restricted by the set $E$. For example, 
$\{1 \mapsto ok,~2 \mapsto ko,~3 \mapsto ok\} \ranres \{ok\}=\{1 \mapsto ok,~3 \mapsto ok\}$.}}
performs a commutation (i.e. changes the closed
  switch),

\item \textbf{Fail} simulates an electrical failure on one of the
  batteries,

\item \textbf{Rep} simulates a maintenance intervention replacing a
  down battery.
\end{itemize}


\begin{figure}
  \centerline{
    \sffamily\scriptsize
      \begin{tabular}{lll}
				$X$ & $\defi$ & $\{H,~Sw,~Bat \}$ \\
				$I$ & $\defi$ & $H \in \{tic, tac\} \wedge Sw\in 1..3 \wedge (Bat\in 1..3\rightarrow \{ok, ko\}) \wedge Bat(Sw)=ok$\\
				$Init$ & $\defi$ & $H,~Sw,~Bat~:=~tac, 1,~\{1 \mapsto ok,~2 \mapsto ok,~3 \mapsto ok\}$ \\
				$Tic$ & $\defi$ & $H=tac \Rightarrow H := tic$ \\
				$Com$ & $\defi$ & $\card (Bat \ranres$
				  $\{ok\})>1 \wedge H=tic \Rightarrow $ \\
				  & &\hspace{0.25cm}%
				  $@ns.(ns \in 1..3 \wedge Bat(ns)=ok \wedge ns \neq Sw \Rightarrow H, Sw := tac, ns)$ \\
				$Fail$ & $\defi$ & $\card (Bat \ranres \{ok\})>1 \Rightarrow$ \\
				  & &\hspace{0.25cm}$@nb.(nb \in 1..3 \wedge nb \in \dom (Bat \ranres \{ok\}) \Rightarrow$ \\
          & & \hspace{0.7cm} $(nb=Sw \Rightarrow @ns.(ns \in 1..3 \wedge ns \neq Sw \wedge Bat(ns)=ok 
          \Rightarrow Sw,Bat(nb) := ns,ko))$ \\
          & & \hspace{0.55cm} $[](nb\neq Sw \Rightarrow Bat(nb):=ko))$ \\
%
%
				$Rep$ & $\defi$ &  $@nb.(nb \in 1..3 \wedge nb \in \dom (Bat \ranres \{ko\}) 
				  \Rightarrow Bat(nb):=ok)$ 
			\end{tabular}
	}
	\caption{\B Specification of the Electrical System\label{fig:BSpecElecSystem}}
\end{figure}

%% file: sec4-SyntacticAbstraction.tex
\section{Syntactic Abstraction}
\label{sec:SyntacticAbstraction}

We define
in this paper a syntactical abstraction method
that applies to \B models. Similar
rules could be adapted for more generic formalisms such as pre-post
models or transition systems. 

Our intention is to obtain an abstract model \abs of a model \fm by
observing only a subset $X_\abs$ of the state variables $X_\fm$ of
\fm.  For instance, to test the electrical system in the particular
cases where two batteries are down, we 
observe only the variable $\textit{Bat}$.  But to preserve the
behaviors of \fm related to the variables of $X_\abs$, we also 
keep in \abs the variables used to assign the observed variables or to
define the conditions under which they are assigned.

%

We first present two methods to compute a set of abstract variables
according to a set of observed variables. Using these variables we
define a predicate and substitution transformation function. Then we
describe how to compute an abstraction of a \B event model \fm. The
abstraction is a bisimulation of \fm when the abstract variables were
computed according to the second method.
We also prove that if they were computed according to the first
method, the abstraction is a simulation of \fm.

\subsection{Choosing the Abstract Variables}
\label{subsec:choixAbstractionVariables}
As proposed in~\cite{SlicingDeModelFormelPourVerif}, we distinguish
between the \emph{observed} variables and the \emph{abstract} ones. A
set $X_\abs$ of \textit{abstract variables} is the union of a set
of \textit{observed variables} with a set of \textit{relevant
  variables}. The \textit{Observed variables} are the ones used by the
tester in a test purpose, while the \textit{relevant variables} are
the ones used to describe the evolutions of the observed variables.
More precisely, the relevant variables are the ones used to assign
an observed variable (\textit{data-flow dependency}), augmented with
the variables used to express when such an assignment occurs
(\textit{control-flow dependency}).

A naive method to define $X_\abs$ is to syntactically collect
the 
variables that are either on the right side or in the guard of the
assignment of an observed variable. But this method will in most cases 
select a very large amount of variables, mainly because of the
guard. For instance, if $x$ is the observed variable, then $y$ is not
relevant in ${ (y \Rightarrow x,z:=E,F) [] (\neg y \Rightarrow x:=E)
}$.  A similar weakness goes for the unbounded
non-deterministic choice ${@z.(P \Rightarrow S)}$.


Hence our contribution consists of two methods for identifying the
relevant variables. The first one only considers the data-flow
dependency. It is efficient, but may select a set too small of
relevant variables, resulting in a set with too many behaviors in the
abstracted model.  The second one uses both data and control flow
dependencies, but requires a predicate simplification to restrict the
size of $X_\abs$. It produces abstract models that have the same set
of behaviors as the original model, w.r.t. the abstract variables.
This second method may select a set with too many relevant
variables 
because predicate simplification is an undecidable problem.


\subsubsection{Proposition 1: Data-Flow Dependency Only}
This first method considers as relevant only the variables that appear
on the right side of an assignment symbol to an abstract variable.
Starting from the set of observed variables, the set of all abstract
variables is computed as the least fix-point when adding the relevant
variables.  For instance, the set of relevant variables of the
electrical system is empty if the set of observed variables is
$\{Bat\}$.  Hence if a test purpose is only based on $Bat$, then
$X_\abs=\{Bat\}$.
A drawback of this method is that it can introduce in \abs new
execution traces w.r.t. \fm. Indeed, it may weaken the guards of some
of the events, that would thus become enabled more often.

\subsubsection{Proposition 2: Data-Flow and Control-Flow Dependencies}
This second method first computes a predicate characterizing a
condition under which an abstract variable is modified, then
simplifies it, and finally considers all its free variables as
relevant.  
%
We express by means of formula~\ref{eqn7} the modifications really
performed by a substitution $S$ on a set $X_\abs$:
\begin{small}
\begin{eqnarray}
\label{eqn7}
Mod_{X_\abs}(S) \defi Prd_{X_\abs}(S) \wedge (\bigvee_{x\in X_\abs} x\neq x').
\end{eqnarray}\vspace*{-10pt}
\end{small}

Our intention is that the predicate, that defines the condition under
which an abstract variable is modified, only involves the variables
really required to modify it. Hence primed variables are not
quantified, but are allowed to be free.
%
%
%
%
For instance, consider $X_\abs=\{x\}$ and the substitution ${x\!:=\!y []
  (z\!>\!0 \Rightarrow x\!:=\!w) [] v\!:=\!3}$.  The predicate has to be in the
shape of: $x'=y \vee (z>0 \wedge x'=w)$, where the variables $y$, $w$
and $z$ are relevant whereas $v$ is not.

\ifLNCSVersion{}{
  The $Mod_{X_\abs}$ predicate can also be defined by induction on the
  primitive substitutions, as described in
  appendix~\ref{Annexe:ModXInductiveDefinition}.
}

%
%


Finally, $X_\abs$ is computed as a least fix-point, by iteratively
incrementing for each event the initial set of observed variables with
the relevant variables.
This process terminates since the set of variables is finite. For
instance, $Mod_{\{Bat\}}$ gives an empty set of relevant variables
when applied to the example, as shown in
Fig.~\ref{fig:SemanticChoiceBat}, while $Mod_{\{H\}}$ gives
$X_\abs=\{Bat,H\}$.

\begin{figure}[ht]
  \scriptsize
  \centerline{
  \begin{tabular}{lcl}
    $Mod_{\{Bat\}}(Init)$&$\Leftrightarrow$&$ Bat=\{1 \mapsto ok,~2 \mapsto ok,~3 \mapsto ok\}$\\
    $Mod_{\{Bat\}}(Tic) $&$\Leftrightarrow$&$ false$ (no assignment of $Bat$)\\
    $Mod_{\{Bat\}}(Com) $&$\Leftrightarrow$&$ false$ (no assignment of $Bat$)\\
    $Mod_{\{Bat\}}(Fail)$&$\Leftrightarrow$&$\card(Bat \ranres \{ok\})>1$\\
                         &                 &$\wedge \exists nb.(nb \in 1..3 \wedge nb \in \dom (Bat \ranres \{ok\}) \wedge Bat'(nb)=ko)$\\
    $Mod_{\{Bat\}}(Rep)$&$\Leftrightarrow$&$ \exists nb.(nb \in 1..3 \wedge nb \in \dom (Bat \ranres \{ko\}) \wedge Bat'(nb)=ok)$
  \end{tabular}
  }
  \normalsize
  \caption{$Mod_{\{Bat\}}$ Computation Applied to the Example}
  \label{fig:SemanticChoiceBat}
\end{figure}

%

\subsection{Predicate Transformation}
\label{subsec:PredTransRules}

Once the set of abstract variables $X_\abs (\subseteq X_\fm)$ is defined, 
we have to describe how to abstract a model according to $X_\abs$. 
We first define the transformation function $T_{X_\abs}(P)$ that 
abstracts a predicate $P$ according to $X_\abs$.
We define $T_X$ on predicates in the 
conjunctive form (see Def.~\ref{DefCNF}) by induction 
with the rules given in Fig.~\ref{fig:PredTransRules}.

An elementary predicate is left unchanged when all the variables used 
in the predicate are considered in the abstraction (see the rule \ref{eq:R3}).
Otherwise, when an expression depends on some variables not kept 
in the abstraction, an elementary predicate is 
undetermined (see the rule \ref{eq:R4}).
As we want to weaken the predicate, we replace an undetermined
elementary predicate by $true$. Consequently, a predicate $P_1 \wedge
P_2$ is transformed into $P_1$ when $P_2$ is undetermined, and a
predicate $P_1 \vee P_2$ is transformed into $true$ when $P_1$ or $P_2$ is
undetermined (see the rules \ref{eq:R5} and \ref{eq:R6}). 
%
Finally, the transformation of a quantified predicate is the
transformation of its body w.r.t. the observed variables, augmented
with the quantified variable (see the rule \ref{eq:R7}).  

\renewcommand\theequation{$R_{\arabic{equation}}$}
\setcounter{equation}{0}

\begin{figure}
  \vspace*{-20pt}
  \begin{scriptsize} 
  \begin{align}
      \label{eq:R3}T_X(E(Y)~r~E(Z)) \defi~& E(Y)~r~E(Z)                  & \textnormal{if }Y \subseteq X \textnormal{ and } Z \subseteq X \\[-2pt]
      \label{eq:R4}T_X(E(Y)~r~E(Z)) \defi~& true                         & \textnormal{if }Y \not\subseteq X \textnormal{ or } Z \not\subseteq X\\[-2pt]
      \label{eq:R5}T_X(P_1 \vee P_2)   \defi~& T_X(P_1) \vee T_X(P_2)          &\\[-2pt]
      \label{eq:R6}T_X(P_1 \wedge P_2) \defi~& T_X(P_1) \wedge T_X(P_2)        &\\[-2pt]
      \label{eq:R7}T_X( \alpha z.P) \defi~& \alpha z.T_{X \cup \{z\}}(P) & 
  \end{align}
  \end{scriptsize}
  \vspace*{-20pt}
%
  \caption{CF Predicate Transformation Rules}
  \label{fig:PredTransRules}
\end{figure}




For example the invariant $I$ of the electrical system is
transformed, according to the single variable $Bat$, into 
$T_{\{Bat\}}(I) = Bat \in 1..3 \rightarrow \{ok, ko\} $
as in Fig.~\ref{fig:ExPredTrans}.

\begin{figure}
	\centerline{\sffamily\scriptsize
      \begin{tabular} {lll}
				& $T_{\{Bat\}}( H\!\in\! \{tic, tac\} \wedge Sw \!\in\! 1..3
				\wedge
				Bat \!\in\! 1..3 \rightarrow \{ok, ko\}  
				\wedge
        Bat(Sw)\!=\!ok) $ & \\
%
				=		&	$\begin{array}{ll}&T_{\{Bat\}}( H\!\in\!\{tic, tac\}) \wedge T_{\{Bat\}}(Sw\!\in\!1..3 
				) \\
				\wedge&
				T_{\{Bat\}}(Bat\!\in\!1..3 \rightarrow \{ok, ko\}) 
        \wedge T_{\{Bat\}}(Bat(Sw)=ok)
        \end{array}$ & applying \ref{eq:R6} \\
				=		&	$Bat \in 1..3 \rightarrow \{ok, ko\}
        $
        & applying \ref{eq:R3} and \ref{eq:R4} \\
%
%
			\end{tabular}
  }
	\caption{Example of Predicate Transformation\label{fig:ExPredTrans}}
\end{figure}

\begin{property}
  \label{property1} Let $P$ be a \CNF predicate in $\Pred$ and let $X$
  be a set of variables. $P \Rightarrow T_X(P)$ is valid.
\end{property}

\begin{proof}
  As we said before, $T_X(P)$ is weaker than $P$. Indeed, for any
  predicate $P$ in \CNF there exist $p_1$ and $p_2$ such that $P=p_1 \wedge
  p_2$ and such that it is transformed either into $p_1 \wedge p_2$, or into
  $p_1$, or into $p_2$, or into $true$, by application of the transformation
  rules $R_i$.  
  For any disjunctive predicate $P$ there exist $p_1$ and $p_2$ such that
  $P=p_1 \vee p_2$ and $p_1 \vee p_2$ is
  transformed either into $p_1 \vee p_2$ or into $true$.
\end{proof}

\subsection{Substitution Transformation}
\label{subsec:SubTransRules}

%

The abstraction of substitutions is defined through cases in
Fig.~\ref{fig:SubTransRules} on the primitive forms of substitutions.
Intuitively, 
any assignment $x:=E$ is preserved into the transformed model if and
only if $x$ is an abstract variable. According to both of the two
methods described in sec.~\ref{subsec:choixAbstractionVariables}, if
$x$ is an abstract variable, then so are all the variables in $E$.
Therefore, in rules~\ref{eq:R8} to~\ref{eq:R14}, we do not transform
the expressions $E$ and $F$.

A substitution is abstracted by $skip$ 
when it does not
modify any variable from $X$ (see rules~\ref{eq:R8}, \ref{eq:R10}, 
\ref{eq:R11} and \ref{eq:R12} in which $y~:=~F$ is abstracted by $skip$). 
The assignment of a variable $x$ is left unchanged if $x$ is an abstract 
variable (see rules~\ref{eq:R9}, \ref{eq:R12}, 
\ref{eq:R14}).  
%
The transformation of a guarded substitution $S$ is such that $T_X(S)$
is enabled at least as often as $S$, since $T_X(P)$ is weaker than $P$
from Prop.~\ref{property1} (see rule~\ref{eq:R16}).
%
The bounded non deterministic choice $S_1\:[]\:S_2$ becomes
a bounded non deterministic choice between the abstraction of $S_1$ and 
$S_2$ (see rule \ref{eq:R18}). 
The quantified substitution is transformed by inserting the bound
variable into the set of abstract variables (see rule \ref{eq:R20}).


\begin{figure}
  \vspace*{-20pt}
  \begin{scriptsize} 
  \begin{align}
    \label{eq:R8}  T_X(x~:=~E)          \defi~& skip   & \textnormal{if } x \notin X \\[-2pt]
    \label{eq:R9}  T_X(x~:=~E)          \defi~& x~:=~E & \textnormal{if } x \in X \\
    \label{eq:R10} T_X(skip)            \defi~& skip   &  \\
    \label{eq:R11} T_X(x,~y:=~E,~F)     \defi~& skip   & \textnormal{if } x \notin X \textnormal{ and } y \notin X\\[-2pt]
    \label{eq:R12} T_X(x,~y:=~E,~F)     \defi~& x~:=~E & \textnormal{if } x \in X \textnormal{ and } y \notin X\\[-2pt]
    \label{eq:R14} T_X(x,~y:=~E,~F)     \defi~& x,~y:=~E,~F          & \textnormal{if } x \in X \textnormal{ and } y \in X\\
    \label{eq:R16} T_X(P \Rightarrow S) \defi~& T_X(P) \Rightarrow T_X(S) & \\ 
    \label{eq:R18} T_X(S_1[]S_2)           \defi~& T_X(S_1)[]T_X(S_2)      & \\ 
    \label{eq:R20} T_X(@z.S)            \defi~& @z.T_{X\cup\{z\}}(S) & 
  \end{align}
  \end{scriptsize}
  \vspace*{-20pt}
%
  \caption{Primitive Substitution Transformation Rules\label{fig:SubTransRules}}
\end{figure}

\subsection{B Event System Transformation }


According to the predicate and substitution transformation functions
(see figure~\ref{fig:PredTransRules} and
figure~\ref{fig:SubTransRules}), we define the transformation of a \B
event model according to a set of abstract variables
(section~\ref{subsec:choixAbstractionVariables}) in
Def.~\ref{DefBeventSystemAbstraction}.  This transformation translates
a correct model \fm into a model \abs that simulates \fm
(Sec.~\ref{sub:sec:correctness}).  The electrical system is
transformed as shown in Fig.~\ref{fig:AbsElecSystem} for the set of
abstract variables $\{Bat\}$.

\begin{definition}[\B Event System Transformation] 
\label{DefBeventSystemAbstraction}
Let $X_\abs$ be a set of abstract variables, defined as in
Sec.~\ref{subsec:choixAbstractionVariables} from a set of observed
variables $X$ with $X \subseteq X_\fm$.  A correct \B event system
\fm=$\BESM$ is abstracted as the \B event system
${\abs=\BESA}$ as follows:
  \begin{itemize}
    \item $X_\abs \subseteq X_\fm$, the set of abstract variables is a subset of the state variables,
    \item $I_\abs=T_{X_\abs}(I_\fm)$, the invariant is transformed,
    \item $Init_\abs=T_{X_\abs}(Init_\fm)$, the initialization is
      transformed,
    \item to each event $ev \defi S_\fm$ in $\Ev_\fm$ is associated $ev \defi
      T_{X_\abs}(S_\fm)$ in $\Ev_\abs$.
  \end{itemize}
\end{definition}


\subsection{Correctness}
\label{sub:sec:correctness}

When the set of abstract variables $X_\abs$ preserve both the data and
control flows as defined in
Sec.~\ref{subsec:choixAbstractionVariables} (Proposition 2), the
transition relation, restricted to $X_\abs$, is preserved, as proved
\ifLNCSVersion{}{(see appendix~\ref{Annexe:TracesEquality:proof})} by
theorem~\ref{theorem3}. \abs and \fm have an equivalent before-after
relation $Prd_{X_\abs}$, therefore they are bisimilar.  Hence when a
CTL* property is verified on \abs it holds on \fm and test cases
generated from \abs can always be  instantiated on \fm.
 

\begin{theorem}
  \label{theorem3}
  Let $S$ be a substitution. Let $X$ be a set of abstract variables composed of any free
  variable of $Mod_{X}(S)$, 
  we have $Prd_{X}(S) \Leftrightarrow Prd_{X}(T_{X}(S))$.
\end{theorem}


With the method defined in Sec.~\ref{subsec:choixAbstractionVariables}
by Proposition~1, \abs is a simulation of \fm.  The \B refinement
relation (see Def.~\ref{DefBeventSystemRefinement}) is proven
in~\cite{bjk00} to be a simulation: \abs simulates \fm by a
$\tau$-simulation. $\tau$ is a silent action corresponding in our case
to an event reduced to $skip$ or to $P \Rightarrow skip$.
Theorems~\ref{theorem1} and~\ref{theorem2} establish that \fm refines
\abs, and thus that \abs simulates \fm.
The safety properties are preserved, but some tests generated from
\abs might be impossible to instantiate on \fm.

\begin{theorem}
  \label{theorem1} Let $I$ be a \CNF invariant of a correct \B event
  system, let $S$ be a substitution and let $X$ be a set of abstract
  variables.  The transformation rules \ref{eq:R8} to \ref{eq:R20} are such that
  $S$ refines $T_X(S)$ according to the invariant $I$.
\end{theorem}

\begin{theorem}
  \label{theorem2} Let $X$ be a set of abstract variables defined as
  in Proposition 1.
  Let $T_X$ be the transformation defined in
  Fig.~\ref{fig:SubTransRules}, and let \abs be an abstraction of an event system \fm
  defined according to Def.~\ref{DefBeventSystemAbstraction}.  \abs is
  refined by \fm in the sense of Def.~\ref{DefBeventSystemRefinement}.
\end{theorem}

Theorem~\ref{theorem1} establishes that any substitution $S$ refines
its transformation $T_X(S)$ for a given set of abstract variables
$X$. \ifLNCSVersion{}{The proof is given in Appendix~\ref{theorem1:proof}.}
Theorem~\ref{theorem2} establishes that a \B event system \fm refines
the \B abstract system obtained according to
Def.~\ref{DefBeventSystemAbstraction} by applying to \fm the
transformation rules of Fig.~\ref{fig:PredTransRules} and
Fig.~\ref{fig:SubTransRules}.

\begin{proof}[of theorem~\ref{theorem2}]
  This is a direct consequence of theorem~\ref{theorem1} and
  Def.~\ref{DefBeventSystemAbstraction} since the substitution
  $Init_\abs \defi T_{X}(Init_\fm)$ is refined by $Init_{\fm}$, and
  that for any event $ev \defi S_{\fm}$, the substitution $S_\abs
  \defi T_{X}(S_\fm)$ is refined by $S_\fm$.
\end{proof}

\begin{figure}
  \centerline{\sffamily\scriptsize
    \begin{tabular}{lll} 
      $X$ & $\defi$ & $\{Bat \}$ \\
      $I$ & $\defi$ & $Bat \in 1..3 \rightarrow \{ok, ko \}$ \\
      $Init$ & $\defi$ & $Bat~:=~\{1 \mapsto ok,~2\mapsto ok,~3 \mapsto ok\}$ \\
      $Tic$ & $\defi$ & $skip$ \\
      $Com$ & $\defi$ &  $\card (Bat \ranres \{ok\})>1 \Rightarrow
        @ns.(ns \in 1..3 \wedge Bat(ns)=ok \Rightarrow skip)$ \\
      $Fail$ & $\defi$ & $\card (Bat \ranres \{ok\})>1 \Rightarrow$ \\
      & & \hspace{0.2cm} $@nb.(nb \in 1..3 \wedge nb \in \dom (Bat \ranres \{ok\}) 
        \Rightarrow Bat(nb) := ko)$ \\
      $Rep$ &  $\defi$ & $@nb.(nb \in 1..3 \wedge nb \in \dom (Bat \ranres \{ko\}) \Rightarrow
      Bat(nb) := ok)$ 
    \end{tabular}
  }
  \caption{\B Syntactically Abstracted Specification of the Electrical System}
  \label{fig:AbsElecSystem}
\end{figure}

%% file: sec5-AbstractionProcess.tex
\newcommand{\SP}{\textsf{SP}\xspace} 
\newcommand{\AST}{\textsf{AST}\xspace} 
\newcommand{\IT}{\textsf{IT}\xspace} 

\section{Application of the Method to a Testing Process}
\label{sec:process}
We show in this section how to use the syntactic abstraction in a
model-based testing approach.

\subsection{Test Generation from an Abstraction}
\begin{figure}[th]
\begin{center}
\includegraphics[scale=0.35]{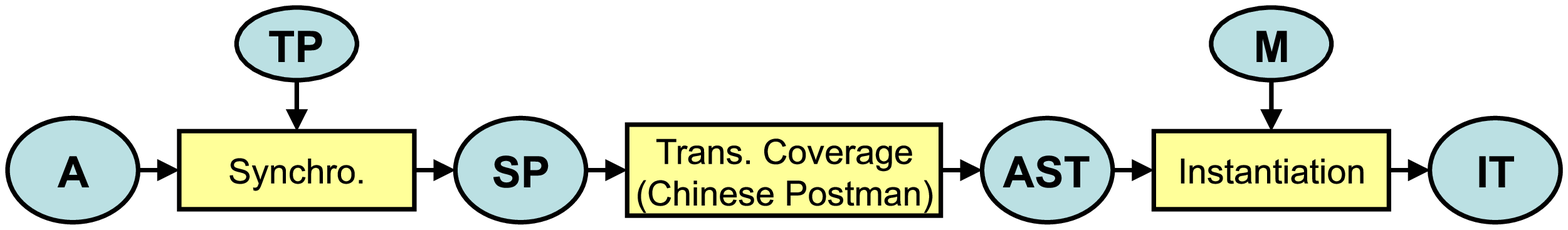} 
\caption{Generating Tests from Test Purpose by Abstraction}
\label{fig:genTestByAbs}
\end{center}
\end{figure}

We have described in~\cite{bbjm-amost10} a model-based testing process using
an abstraction as input. It can be summarized as follows. A validation
engineer describes by means of a handwritten test purpose \tp how he
intends to test the system, according to his know-how.  We have
proposed in~\cite{jmt-ast08-langage} a language based on regular
expressions, to describe a \tp as a sequence of actions to fire and
states to reach (targeted by these actions). The actions can be
explicitly called in the shape of event names, or left unspecified by
the use of a generic name. The unspecified calls then have to be
replaced with explicit event names. However, a combinatorial explosion
problem occurs, when searching in a concrete model for the possible
replacements that lead to the target states. This leads us to use
abstractions instead of concrete models. Figure~\ref{fig:genTestByAbs}
shows our approach.

We perform a synchronized product between an abstraction \abs and the
automaton of a \tp. This results in a model \SP whose executions are
the executions of \abs that match the \tp. An
implementation~\cite{Thi03} of the Chinese Postman algorithm is
applied to \SP to cover its transitions. The result is a set of
abstract symbolic tests \AST. These tests are instantiated from \fm as
a set \IT of instantiated tests.

\subsection{Abstraction Computation}
We show in this section two ways of producing an abstraction \abs that
can be used as an input of the process of
Fig.~\ref{fig:genTestByAbs}. The syntactic abstraction of
Sec.~\ref{sec:SyntacticAbstraction} is used in one of these two ways.

In order to compute the synchronized product of an abstraction \abs
with the automaton of a \tp, we compute the semantics of \abs as a
labelled transition system. We use \genesyst~\cite{genesystZB05} for
that purpose.
This tool computes a semantic abstraction of a B model in the shape of
a symbolic labelled transition system.
The semantic abstraction relies on feasibility proofs of the
transitions between two symbolic states. \genesyst generates proof
obligations (\PO{}s) for each of the potential transitions between two
symbolic states, and tries to solve them automatically.

The two main drawbacks of this process are its time cost and the
proportion of \PO{}s not automatically solved.  Indeed, each unsolved
\PO results in a transition that is kept in the symbolic labelled
transition system, although it is possibly unfeasible. An abstract
symbolic test going through such a transition may be impossible to
instantiate from the concrete model \fm. By applying a preliminary
phase of syntactic abstraction, we reduce the impact of that problem
by reducing the number and the size of the \PO{}s, since \genesyst
operates on an already abstracted model. For example, no proof
obligation is generated for an event reduced to $skip$ (it becomes a
reflexive transition on any symbolic state).




\begin{figure}
\centerline{
\includegraphics[width=\textwidth]{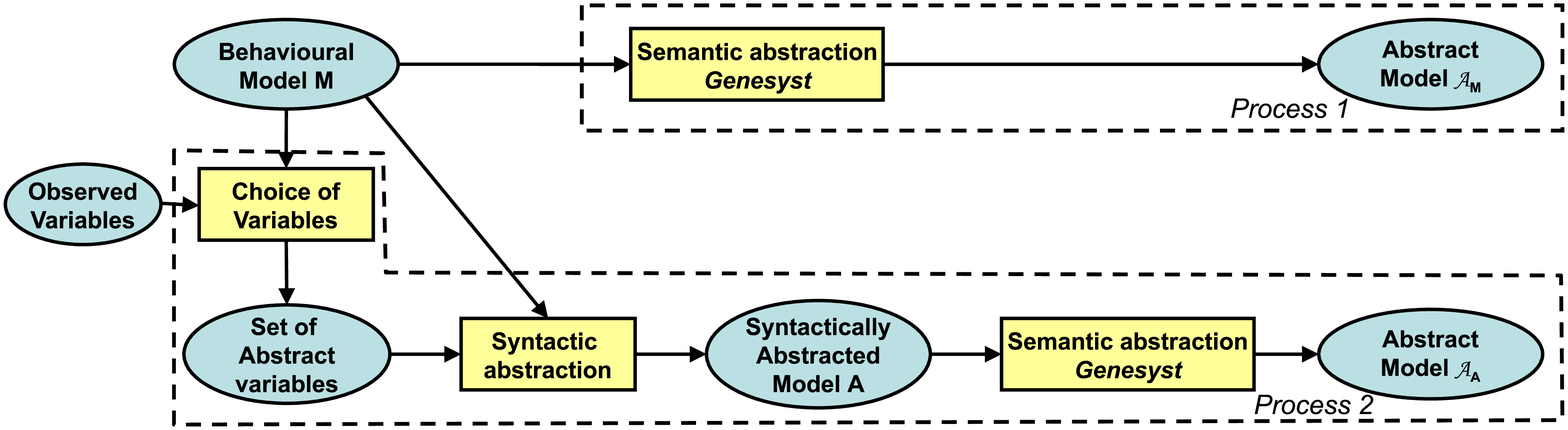}
}
\vspace*{-6pt}
\caption{Abstraction Process}
\vspace*{-10pt}
\label{fig:AbsProc}
\end{figure}

The experimental results presented in Sec.~\ref{sec:expResults}
compare two approaches. The first one (see
Fig.~\ref{fig:AbsProc}/Process~1) is only semantic, while the second
one (see Fig.~\ref{fig:AbsProc}/Process~2) combines a syntactic and a
semantic abstraction.

%% file: sec6-experimentalResults.tex
\section{Experimental Results}
\label{sec:expResults}
We have applied our method to four case studies. They are 
various cases of reactive systems: an automatic conveying system
(Robot~\cite{bbjm09}), a reverse phone book service
(Qui-Donc~\cite{MBTbook06}), the electrical system\footnote{The 100
  lines length of the model, in Table \ref{tab:mesureModels}, refer to
  a ``verbose'' version of the model, much more readable than our
  version of
  Fig.~\ref{fig:BSpecElecSystem}.
} (Electr.
) and an electronic purse
(DeMoney~\cite{specDemoney-tl-02}). Each one is abstracted w.r.t. two
sets of abstract variables. These sets have been computed according to
Proposition~1 of Sec.~\ref{subsec:choixAbstractionVariables}.  We also
have tried to compute the abstract variables according to
Proposition~2, but all the variables have been computed as
\emph{abstract} in three case studies. Only for the electrical system
the set of abstract variables was the same as with Proposition~1.
These case studies reveal a limit in the application of Proposition~2.



In Sec.~\ref{sec:SynAbs} we present an experimental evaluation of the
syntactic abstraction.  Then, in Sec.~\ref{sec:SemAbs} we
compare $\AM$ with $\AA$ respectively computed by the semantic
abstraction process or by its combination with the syntactic one.

\subsection{Impact of the Syntactic Abstraction on Models}
\label{sec:SynAbs}

Table~\ref{tab:mesureModels} indicates the size of the case studies
and the syntactically abstracted models. 
The Symbols ``$\sharp$'', ``Ev.'', ``Var.'' and ``Pot.''
respectively stand for \textit{number of}, \textit{Events},
\textit{Variables} and \textit{Potential}.  For example the Robot,
defined by 9 events and 6 variables is abstracted w.r.t. two sets of
respectively 3 and 4 abstract variables.


\begin{table}
	\centerline{\scriptsize
	\begin{tabular}{|l|c||*{2}{c|}c||*{4}{c|}}
		\hline
		\multirow{2}*{Case Study} & \multirow{2}*{$\sharp$Ev.} & \multicolumn{3}{c||}{Model \fm} & \multicolumn{4}{c|}{Syntactically abstracted model \abs} \\ 
		\cline{3-9}	
		& & $\sharp$Var. & $\sharp$B lines & $\sharp$Pot. states & $\sharp$Var. & $\sharp$B lines & $\sharp$Pot. states & $\sharp$Symb. states \\
		\hline
		\multirow{2}*{Robot} & \multirow{2}*{9} & \multirow{2}*{6} & \multirow{2}*{100} & \multirow{2}*{384} & 3 & 90 & 48 & 6 \\
		\cline{6-9}
		& & & & & 4 & 90 & 144 & 8 \\
		\hline
		\multirow{2}*{QuiDonc} & \multirow{2}*{4} & \multirow{2}*{3} & \multirow{2}*{170} & \multirow{2}*{13} & 2 & 160 & 16 & 5 \\
		\cline{6-9}
		& & & & & 2 & 160 & 16 & 6 \\
		\hline
		\multirow{2}*{Electr.} & \multirow{2}*{4} & \multirow{2}*{3} & \multirow{2}*{100} & \multirow{2}*{36} & 1 & 50 & 5 & 2 \\
		\cline{6-9}
		& & & & & 1 & 40 & 2 & 2 \\
		\hline
		\multirow{2}*{DeMoney} & \multirow{2}*{11} & \multirow{2}*{9} & \multirow{2}*{330} & \multirow{2}*{$10^{30}$} & 1 & 140 & 65536 & 3 \\
		\cline{6-9}
		& & & & & 2 & 180 & 7 & 4 \\
		\hline
	\end{tabular}
  } 
	\caption{Size of the Case Studies and of their Syntactical Abstractions}
	\label{tab:mesureModels}
\end{table}

A direct observable result of the syntactic abstraction is a reduction
of the number of potential states of the model.  Also notice that the
simplification reduces from 10\% up to 50\% the number of lines of the
model.

\subsection{Comparison of the Abstraction Processes 1 and 2}
\label{sec:SemAbs}

\begin{table}
	\centerline{\tiny
	\begin{tabular}{|l|*{13}{c|}}
		\hline
		\multirow{2}*{Case} & \multicolumn{6}{c|}{Process 1 : $\AM$} & \multicolumn{6}{c|}{Process 2 : $\AA$} & \multirow{2}*{Traces} \\ 
		\cline{2-13}
		\multirow{2}*{study} & 
                \multirow{2}*{$\sharp$Trans.} & $\sharp$Unau. &
                \multirow{2}*{$\sharp$\PO} & Time & $\sharp$Inst./ &
                Trans. Cover. & \multirow{2}*{$\sharp$Trans.} &
                $\sharp$Unau. & \multirow{2}*{$\sharp$\PO} & Time &
                $\sharp$Inst./ & Trans. Cover. & \multirow{2}*{inclusion} \\
		& & Trans. & & (s) & $\sharp$Tests. & of $\AM$ & &
                Trans. & & (s) & $\sharp$Tests. & of $\AA$ & \\
		\hline
		\multirow{2}*{Robot} & 42 & 5 & 263 & 64 & 4/11
                & 29/37 (78\%) & 36 & 0  & 143 & 35 & 7/11
                & 31/36 (86\%) &  $\A_\abs \subset \A_\fm$ \\
		\cline{2-14}
		& 51 & 0 & 402 & 76 & 4/23 & 35/51 (68\%) &
                50 & 0 & 242 & 49 & 8/23 & 38/50 (76\%) & $\A_\abs \subset \A_\fm$ \\
		\hline
		Qui- & 20 & 2 & 71 & 19 & 9/11
                & 12/18 (66\%) & 25 & 7 & 89 & 21 & 6/11
                & 11/18 (61\%) &  $\A_\abs \nsubseteq \A_\fm$ \\
		\cline{2-14}
		Donc & 25 & 2 & 89 & 21 & 4/10 & 6/23 (26\%) &
                29 & 6 & 103 & 23 & 4/10 & 6/23 (26\%) & $\A_\abs \nsubseteq \A_\fm$ \\
		\hline
		\multirow{2}*{Electr.} & 13 & 5 & 26 & 7 & 2/2
                & 8/8 (100\%) & 13 & 5 & 16 & 5 & 2/2
                & 8/8 (100\%) & $\A_\abs = \A_\fm$ \\
		\cline{2-14}
		& 7 & 0 & 21 & 5 & 3/3 & 7/7 (100\%) & 7 &
                0 & 9 & 2 & 3/3 & 7/7 (100\%) & $\A_\abs = \A_\fm$ \\
		\hline
		De- & 38 & 5 & 116 & 189 &
                17/18 & 25/33 (76\%) & 38 & 5 & 68 & 38 & 17/18
                & 25/33 (76\%) & $\A_\abs \subset \A_\fm$ \\
		\cline{2-14}
		Money & 53 & 0 & 290 & 172 & 22/38 & 30/53 (56\%)
                & 50 & 0 & 130 & 65 & 20/35 & 26/50 (52\%) & $\A_\abs \subset \A_\fm$ \\
		\hline
	\end{tabular}
	}
	\caption{Comparison of the semantic and syntactic/semantic abstraction processes}\vspace*{-10pt}
	\label{tab:mesureAbstract}
\end{table}

Table~\ref{tab:mesureAbstract} compares the abstractions computed
either directly from the behavioral models (see process~1 in
Fig.~\ref{fig:AbsProc}), or from their syntactic abstractions (see
process~2 in Fig.~\ref{fig:AbsProc}). The abbreviations 
``Trans.'', ``Unau.'', ``Inst.'' and ``Cover.'' stand respectively for
\emph{transitions}, \emph{unauthorized},
\emph{instantiated} and \emph{coverage}.

We see on our examples that there is between $1.8$ and $2.3$ fewer
\PO{}s to compute with process~2 than with process~1, except for the
Qui-Donc.  The semantic abstraction computation in process~2 takes
from twice up to five times less time than in process~1, where no
previous syntactic abstraction have been performed. For the Qui-Donc,
the syntactical abstraction has too much over-approximated the initial
model, which explains the augmentation of the \PO{}s w.r.t. the
process~1.  Finally, there are four cases out of eight where the
abstraction $\AA$ is more precise than $\AM$ in the sense that it has
less transitions, due to the reduction of the number of unproved
\PO{}s.  In these four cases, the set of traces of $\AA$ is included
in the set of traces of $\AM$.  In the case of the electrical system,
the set of traces are equal.
%
%
In the Qui-Donc case, the traces cannot be compared.
The simplification by the syntactic abstraction of the events and of
the invariant makes that $\AA$ may contain more transitions (thus more
traces) than $\AM$. But the number and the difficulty of the \PO{}s is
greater to get $\AM$ than to get $\AA$, so that proof failures may
occur more often with $\AM$. As a result, $\AM$ can also contain 
transitions that are not in $\AA$.


As for the ratios of tests instantiated and of transitions covered of
the abstraction, we observe their stability with or without syntactic
abstraction. Although the ratios are a bit better (or equal) for the
Robot and the Electrical System, and a bit worse for Qui-Donc and
Demoney, they are mainly very close to each other. But, due to the
reduction of the number of \PO{}s, the time to obtain these comparable
results is improved with process~2, i.e. when there is a preliminary
syntactic abstraction phase. Again, this is not true for the Qui-Donc
since on the contrary, its number of \PO{}s has increased.

Finally, the method had no interest with the Qui-Donc, which was the
smallest example. But, as shown by DeMoney, its efficiency in terms of
gain of the abstraction computation time, of reduction of the number
of unproved \PO{}s and of precision of the abstraction, grows with the
size of the examples.

%% file: sec7-conclusion.tex
\section{Conclusion, Related Works and Further works}
\label{sec:conclusion}
We have presented in the \B framework a method for abstracting an
event system by elimination of some state variables. In this context, 
we have proposed two methods to compute the set of variables 
kept in the abstraction according to the set of observed variables. 
We have proved that when using the first method, the generated abstraction 
simulates the concrete model, while when using the second method, the 
generated abstraction bi-simulates the concrete model.
This is useful for verifying safety properties and generating tests.

In the context of test generation, our method consists in initializing
the test generation process from event \B model described in~\cite{bbjm-amost10}, 
by a syntactic abstraction. Since the syntactic abstraction reduces the size 
of the model, the main advantage of this method 
is that it reduces the set of uninstantiable tests, by reducing the level 
of abstraction (reduces the number of \PO generated and
facilitates the proof of the remaining \PO). Moreover, this results in a 
gain of computation time.
We believe that the bigger the ratio of the number of state variables
to the number of observed variables is, the bigger the gain is. This
conjecture needs to be confirmed by experiments on industrial size
applications.



Many other works define model abstraction methods to verify properties
or to generate tests. The method of~\cite{FriedmanHNS02} uses an
extension of the model-checker Mur$\phi$ to compute tests from
projected state coverage criteria that eliminate some state variables
and project others on abstract domains.  In~\cite{DickF93}, an
abstraction is computed by partition analysis of a state-based
specification, based on the pre and post conditions of the
operations. Constraint solving techniques are used.  The methods
of~\cite{GS97,BLO98,CU98} use theorem proving to compute the abstract
model, which is defined over boolean variables that correspond to a
set of \emph{a priori} fixed predicates. In contrast, our method
first introduces a syntactical abstraction computation from a set of
observed variables, and further abstracts it by theorem proving.
\cite{CABN97} also performs a syntactic transformation, but requires
the use of a constraint solver during a model checking process.


Other automatic abstraction methods~\cite{CGL94} are limited to finite
state systems. The deductive model checking algorithm of ~\cite{SUM99}
produces an abstraction w.r.t. a LTL property by an iterative
refinement process that requires human expertise. Our method can
handle infinite state space specifications.
The paper~\cite{NK00} presents a syntactic abstraction method for
guarded command programs based on assignment substitution.
The method is sound and complete for programs without unbounded non
determinism. However, the method is iterative and does not terminate
in the general case. It requires the user to give an upper-bound of
the number of iterations.
The paper also presents an extension for unbounded non deterministic
programs that is sound but not complete, due to an exponential number
of predicates generated at each iteration step.
In contrast, our syntactic method is iterative on the syntactic
structure of the specifications. It is sound but not complete.  It
handles unbounded non deterministic specifications with no need for
other iterative process and always terminates.  Above all, our method
does not compute any weakest precondition whereas the approach
in~\cite{NK00} does, which possibly introduces infinitely many new
predicates.


The syntactic method that we have presented is correct, but, in the
case of Proposition~1, may sometimes produce inaccurate
over-approximations due to a too strong abstraction (see for example
the experiments on the Qui-Donc). Proposition~2 produces a
bisimulation, but may leave the initial model unchanged, i.e. not
abstracted, if all the variables are computed as abstract. We have to
find a compromise between the two propositions, that would reduce the
number of abstract variables, but that would keep at least partially
the control structure of the operations.  Also, we think that rules
could be improved to get a finer approximation.
For instance, improving the rules is possible when the invariant
contains an equivalence such as $x=c \Leftrightarrow y=c'$.  If 
$y$ is an eliminated variable and $x$ an observed one, we could
substitute all the occurrences of the elementary predicate $y=c'$ with
$x=c$.
This would preserve the property in the syntactic
abstraction $\AA$, so that the following semantic abstraction would be
more accurate. Such rules should prevent the addition of transitions
in the Qui-Donc abstraction $\AA$ w.r.t. $\AM$.


We think that extending the test generation method introduced
in~\cite{bbjm-amost10} by using a combination of syntactic and semantic
abstractions will improve the method, since the abstraction is more
accurate if there are less unproved \PO{}s. 

 
%

%% file: Annexe-ModX.tex
\section{Inductive Definition of $Mod_X$}

The $Mod_{X}$ predicate can be defined by induction through primitive substitutions,
as described in Table~\ref{table:ModPredicateOnPrimitives}.
Intuitively, an assignment $x:=E$ is associated to $false$ if and only if $x$ is not in ${X}$ or $x$ already has the same value as $E$. 
Other assignment cases are just some generalizations. 
This implements the data-flow dependency. 
For control flow dependency, a non-deterministic choice is an union between control-flow branches, thus a disjunction between predicates, and 
a guarded substitution $P\Rightarrow S$ is associated to the whole condition $P$ augmented with the result of the analysis of $S$. Once this predicate is expressed,  
it needs to be logically simplified.

\label{Annexe:ModXInductiveDefinition}

\begin{table}[ht]
  \centerline{\sffamily\scriptsize
    \begin{tabular}{|l@{~~}c@{~~}l@{~~~~~}l|}
      \hline
      \bf Substitution                      && \bf Modification Predicate &\bf Condition \\ 
      \hline
      \hline
      $Mod_X (x  :=  E)$          &$\defi$& $false$ & $x \notin X$\\ 
      $Mod_X (x  :=  E)$          &$\defi$& $x'=E \wedge \bigwedge_{z \in X-\{x\}} (z'=z) \wedge x \neq x'$
         &$x \in X$\\[4pt] 
%
%
      $Mod_X (x,y  :=  E,F)$      &$\defi$& $false$ & $x \notin X \wedge y \notin X$\\ 
      $Mod_X (x,y  :=  E,F)$      &$\defi$& 
          $x'=E \wedge \bigwedge_{z \in X-\{x\}} (z'=z)
           \wedge x\neq x'$ &$x \in X \wedge y \notin X$\\ 
%
      $Mod_X (x,y  :=  E,F)$      &$\defi$& 
          $x'\!=\!E \wedge y'\!=\!F \wedge \bigwedge_{z \in X-\{x, y\}} (z'\!=\!z)
             \wedge \bigvee_{z\in \{x,y\}}(z \!\neq\! z')
           $ & $x \in X \wedge y \in X$\\[4pt]
      $Mod_X (skip)$              &$\defi$& $false$ &\\ 
      $Mod_X (P \Rightarrow S)$   &$\defi$& $P \wedge\ Mod_X(S)$ &\\ 
      $Mod_X (S_1 ~[]~ S_2)$          &$\defi$& $Mod_X(S_1) \vee Mod_X(S_2)$&\\ 
      $Mod_X (@ z \cdot  S)$      &$\defi$& $\exists (z,z') \cdot Mod_{X\cup \{z\}}(S)$ &\\ 
      \hline
    \end{tabular}
  }
  \caption{$Mod_X(S)$ Predicate Defined through Primitive Substitutions }
  \label{table:ModPredicateOnPrimitives}
\end{table}

\begin{property}
  \label{property:Mod} $Mod_X(S)$ defined in Table~\ref{table:ModPredicateOnPrimitives} satisfy the
  definition in formula~(\ref{eqn7}).
\end{property}

\begin{proof}[of property~\ref{property:Mod}]
  For any case of $S$, we prove that $Mod_X(S)$ defined as in
  Formula~(\ref{eqn7}) replacing $Prd_X(S)$ by its definition given in
  formula~(\ref{eqn6}) and transformed by the formulas~(\ref{eqn1}) to
  (\ref{eqn4}) is equal to its value in
  Table~\ref{table:ModPredicateOnPrimitives}.
\end{proof}

%% file: Annexe-proof_th1.tex
\section{Proof of Theorem~\ref{theorem1}}

\newcommand{\consts}{PC_\abs \land PC_\fm}
\newcommand{\rename}{\textsf{Ren}}
\newcommand{\glue}{\textsf{G}}

\label{theorem1:proof}
\begin{proof}

The refinement theory as defined in \B~\cite{Bbook}, requires that
variable sets from abstraction and variable sets from refinement are
disjoint. If a variable $x$ is preserved through the refinement process, then it has to be renamed, i.e. $x_{\textit{renamed}}$, and associated by a gluing invariant, i.e. $x=x_{\textit{renamed}}$.
In order to prove the correctness of the refinement, we introduce the $\rename()$ function, which renames every variable from a substitution or a predicate. Hence, the invariant $I_\abs$ abstracted from $I_\fm$ and the substitution $S_\abs$ abstracted from any $S_\fm$ are defined as follows:

\small
\noindent~\hfill $I_\abs \bdef \rename(T_X(I_\fm))$ \hfill  $S_\abs \bdef \rename(T_X(S_\fm))$ \hfill ~
\normalsize

To prove that $S_\fm$ is a correct refinement of $S_\abs$, we need to prove (Def.~\ref{DefBeventSystemRefinement}):\vspace*{-6pt}
\small
\begin{equation}
\label{proof:global}\consts \land I_\abs \land I_\fm \land I_\glue \Rightarrow [S_\fm]\neg[S_\abs]\neg(I_\fm \land I_\glue)
\end{equation}
\normalsize
where $I_\glue$ is the gluing invariant $I_\glue \bdef \bigwedge_{x_i\in X} (x_i=\rename(x_i))$.
In order to prove formula~(\ref{proof:global}), it is sufficient to
establish that the following two formulas hold:
\small
\vspace*{-6pt}
\begin{eqnarray}
  \label{proof:ref}\consts \land I_\abs \land I_\fm \land I_\glue \Rightarrow [S_\fm]\neg[S_\abs]\neg I_\fm\\
  \label{proof:abstr}\consts \land I_\abs \land I_\fm \land I_\glue \Rightarrow [S_\fm]\neg[S_\abs]\neg I_\glue 
\vspace*{-10pt}
\end{eqnarray}
\normalsize
Since free variable sets from $I_\abs$ and $I_\fm$ are strictly disjoint, (\ref{proof:ref}) can be rewritten as: 
%
$\consts \land I_\abs \land I_\fm \land I_\glue \Rightarrow [S_\fm] I_\fm$, 
%
%
that holds, since the initial model \fm is correct. 
%
Hence, we only have to establish~(\ref{proof:abstr}) to prove Theorem~\ref{theorem1}.
The proof is by induction on the five primitive forms of substitutions. 
We make a case analysis for each rule in Fig.~\ref{fig:SubTransRules}.
We use Prop.~\ref{property1} of Sec.~\ref{subsec:PredTransRules} and axioms (\ref{eqn1} to \ref{eqn5}) defined in 
Sec.~\ref{sec:preliminaries}.

\newcommand{\hyps}{\textit{Hyps}}
We denote by $\hyps$ the repetitive predicate {\small $\hyps \bdef \consts \land I_\abs \land I_\fm \land I_\glue$}.

\begin{description}\small
\item[Case $S_\fm\bdef x~:=~E$]~
\begin{description}
\item [\textbf{Rule \ref{eq:R8}}] $S_\abs\bdef skip$ \hfill\textit{when $x\not\in X$}\\
  is $\hyps \Rightarrow [x~:=~E]\neg[skip]\neg I_\glue$ valid ?\\
  It is valid, according to (\ref{eqn1}), since $x$ is not free in $I_\glue$.

\item [\textbf{Rule \ref{eq:R9}}] $S_\abs\bdef \rename(x):=\rename(E)$ \hfill\textit{when $x\in X$}\\
  is $\hyps \Rightarrow [x~:=~E]\neg[\rename(x):=\rename(E)]\neg I_\glue$ valid ?\\
  It is valid since Rule \ref{eq:R9} is the identity.
\end{description}

\item[Case $S_\fm\bdef skip$]~
\begin{description}
\item[\textbf{Rule \ref{eq:R10}}] $S_\abs\bdef skip$\\
  $\hyps \Rightarrow [skip]\neg[skip]\neg I_\glue$ is obviously valid according to (\ref{eqn1}).
\end{description}

\item [Case $S_\fm\bdef x,~y:=~E,~F$]~
\begin{description}
\item[\textbf{Rules \ref{eq:R11} to \ref{eq:R14}}] proofs are
  similar to the first case.
\end{description}

\item [Case $S_\fm\bdef P \Rightarrow S$]~
\begin{description}
  
  
\item[Rule \ref{eq:R16}]
 $S_\abs\bdef \rename(T_X(P)) \Rightarrow \rename(T_X(S))$ \\ 
  is $\hyps \Rightarrow [P \Rightarrow S]\neg[\rename(T_X(P)) \Rightarrow \rename(T_X(S))]\neg I_\glue$ valid ?\\
  $\equiv \hyps \Rightarrow P \Rightarrow [S](\rename(T_X(P)) \land \neg [\rename(T_X(S))]\neg I_\glue)$  \hfill-- applying (\ref{eqn2})\\
  $\equiv \left\{\begin{array}{lll}
             &(\textnormal{\ref{eq:R16}}.1)&(\hyps \land P \Rightarrow [S]\rename(T_X(P)))\\
             \land&(\textnormal{\ref{eq:R16}}.2)&(\hyps \land P \Rightarrow [S]\neg [\rename(T_X(S))]\neg I_\glue)
   \end{array}\right.$  \hfill-- applying (\ref{eqn5})\\
  According to Prop~\ref{property1}, $(\textnormal{\ref{eq:R16}}.1)$ holds since $S$ variables are not free in $\rename(T_X(P))$ and since $I_\glue$ is in $\hyps$.
  $(\textnormal{\ref{eq:R16}}.2)$ is valid w.r.t. the induction hypothesis:\\ ${\hyps \!\Rightarrow\! [S]\neg[\rename(T_X(S))]\neg I_\glue}$.

\end{description}

\item [Case $S_\fm \bdef S~{[]}~S'$]~
\begin{description}
   
\item[Rule \ref{eq:R18}] $S_\abs\bdef \rename(T_X(S)) [] \rename(T_X(S'))$ \\
  is $\hyps \Rightarrow [S~[]~S']\neg[\rename(T_X(S)) [] \rename(T_X(S'))]\neg I_\glue$ valid ?\\
  $\equiv \!\hyps\! \Rightarrow [S\,[]\,S'](\neg[\rename(T_X(S))]\neg I_\glue \!\lor\! \neg[\rename(T_X(S'))]\neg I_\glue)$ \hfill-- applying (\ref{eqn3})\\
  $\equiv \!\!\left\{\!\begin{array}{l@{\,}l}
                  &(\hyps \!\Rightarrow\! [S](\neg[\rename(T_X(S))]\neg I_\glue \!\lor\! \neg[\rename(T_X(S'))]\neg I_\glue))\\
            \land &(\hyps \!\Rightarrow\! [S'](\neg[\rename(T_X(S))]\neg I_\glue \!\lor\! \neg[\rename(T_X(S'))]\neg I_\glue))
          \end{array}\right.\!$ \hfill-- applying (\ref{eqn3})\\
  This formula is valid because the two induction hypotheses are valid:
  \begin{enumerate}
  \item $\hyps \Rightarrow [S]\neg[\rename(T_X(S))]\neg I_\glue$,
  \item $\hyps \Rightarrow [S']\neg[\rename(T_X(S'))]\neg I_\glue$.
  \end{enumerate}
\end{description}

\item [Case $S_\fm\bdef @z.S$]~
\begin{description}
  
\item[Rule \ref{eq:R20}] $S_\abs\bdef \rename(@z.T_{X \cup
    \{z\}}(S))$ \\ 
  is $\hyps \Rightarrow [@z.S]\neg[\rename(@z.T_{X \cup \{z\}}(S))]\neg I_\glue$ valid ?\\
  $\equiv \hyps \Rightarrow \forall z.[S] \neg \forall \rename(z).[\rename(T_{X \cup \{z\}}(S))] \neg I_\glue$ \hfill-- applying (\ref{eqn4}) \\
  It is valid since the following formula is implied by the induction hypothesis:
  
  \centerline{$\hyps \Rightarrow \forall z.\exists \rename(z).(z=\rename(z) \land [S] \neg [\rename(T_{X \cup \{z\}}(S))] \neg (I_\glue \land z=\rename(z)))$}

\end{description}

\end{description}

\noindent Hence, Theorem~\ref{theorem1} holds.
\end{proof}

%% file: Annexe-proof2.tex
\section{$Prd_X(M)=Prd_X(T_X(M))$ ?}
\label{Annexe:TracesEquality:proof}

Let $S$ be a substitution. Let $X$ be a set of abstract variables composed of any free
variable of $Mod_{X}(S)$ (see Proposition 2 in Sec.~\ref{subsec:choixAbstractionVariables}).
We propose to prove that the following formula holds: ${Prd_{X}(S) \Leftrightarrow Prd_{X}(T_{X}(S))}$.


Since ${Prd_{X}(S) \defi \neg [S] \neg \bigwedge_{x\in X} x=x'}$ (see
formula~(\ref{eqn6}) in
Sec.~\ref{sec:preliminaries}), we verify it by induction through primitive substitutions
proving that $[S]P \Leftrightarrow [T_{X}(S)] P$ holds 
when $P$ is defined only in terms of abstract variables in $X$.

%

Let $[T_{X}(S)]P \Leftrightarrow [S]P$
be the induction hypothesis:
 \begin{scriptsize} 
  \begin{align*}
    \mathbf{[T_{X}(S)] P}~\Leftrightarrow~&\mathbf{[S]P}  & \textbf{Condition or justification}\\
    \hline
    [skip]P              ~\Leftrightarrow~& [x:=E]P & \textnormal{if } x \notin X \\[-2pt]
    [x~:=~E]P            ~\Leftrightarrow~& [x:=E]P & \textnormal{if } x \in X \\
    [skip]P              ~\Leftrightarrow~& [skip]P &  \\
    [skip]P              ~\Leftrightarrow~& [x,~y:=~E,~F]P & \textnormal{if } x \notin X \textnormal{ and } y \notin X\\[-2pt]
    [x~:=~E]P            ~\Leftrightarrow~& [x,~y:=~E,~F]P & \textnormal{if } x \in X \textnormal{ and } y \notin X\\[-2pt]
    [y~:=~F]P            ~\Leftrightarrow~& [x,~y:=~E,~F]P & \textnormal{if } x \notin X \textnormal{ and } y \in X\\[-2pt]
    [x,~y:=~E,~F]P       ~\Leftrightarrow~& [x,~y:=~E,~F]P & \textnormal{if } x \in X \textnormal{ and } y \in X\\
    T_X(P_1)\Rightarrow[T_X(S)]P~\Leftrightarrow~& P_1\Rightarrow [S]P & \textnormal{since } T_X(P_1) = P_1 \textnormal{ according to } \\
                                          &                & Mod_X(P_1 \Rightarrow S) \textnormal{ definition.} \\
    [T_X(S_1)[]T_X(S_2)]P   ~\Leftrightarrow~& [S_1[]S_2]P       & \textnormal{by induction hypothesis} \\
  [@z.T_{X\cup\{z\}}(S)]P~\Leftrightarrow~& [@z.S]P       & \textnormal{by formula~\ref{eqn5} and induction hypothesis}
  \end{align*}
  \end{scriptsize}
  
Notice that the hypothesis when $P$ is defined only in terms of abstract variables $X$
induces that $[x := E]P = P$ when $x \notin X$ because there is no occurrence of $x$ in $P$.

We can then conclude that the set of behaviors on the set of abstract variables $X$ of an event $ev$ is unchanged when we simplify it by $T_X$.

%

%% file: btap10-jsbm.bbl
\begin{thebibliography}{10}

\bibitem{MBTofReacSys05}
Broy, M., Jonsson, B., Katoen, J.P., Leucker, M., Pretschner, A., eds.:
\newblock Model-Based Testing of Reactive Systems. Volume 3472 of LNCS.
\newblock Springer (2005)

\bibitem{MBTbook06}
Utting, M., Legeard, B.:
\newblock Practical Model-Based Testing - A tools approach.
\newblock Elsevier Science (2006)

\bibitem{ProB}
Leuschel, M., Butler, M.:
\newblock {ProB}: An automated analysis toolset for the {B} method.
\newblock Software Tools for Technology Transfer \textbf{10}(2) (2008)
  185--203

\bibitem{bcdg-B2007}
Bouquet, F., Couchot, J.F., Dadeau, F., Giorgetti, A.:
\newblock Instantiation of parameterized data structures for model-based
  testing.
\newblock In: B'2007, the 7th Int. B Conference. Volume 4355 of LNCS., Springer
  (2007)  96--110

\bibitem{bbjm-amost10}
Bouquet, F., Bu\'e, P.C., Julliand, J., Masson, P.A.:
\newblock Test generation based on abstraction and test purposes to complement
  structural tests.
\newblock In: A-MOST'10, 6th int. Workshop on Advances in Model Based Testing,
  Paris, France (April 2010)

\bibitem{specDemoney-tl-02}
Marlet, R., Mesnil, C.:
\newblock Demoney: A demonstrative electronic purse 
\newblock Technical Report SECSAFE-TL-007, Trusted Logic (2002)

\bibitem{genesystZB05}
Bert, D., Potet, M.L., Stouls, N.:
\newblock {GeneSyst: a Tool to Reason about Behavioral Aspects of B Event
  Specifications}.
\newblock In: ZB'05. Volume 3455 of LNCS. (2005)

\bibitem{Weiser84}
Weiser, M.:
\newblock Program slicing.
\newblock Software Engineering, IEEE Transactions on \textbf{SE-10}(4) (july
  1984)  352--357

\bibitem{CGS09}
Couchot, J.F., Giorgetti, A., Stouls, N.:
\newblock {Graph-based Reduction of Program Verification Conditions}.
\newblock In: {AFM'09}. (2009)

\bibitem{Bbook}
Abrial, J.R.:
\newblock The {B} Book: Assigning Programs to Meanings.
\newblock Cambridge University Press (1996)

\bibitem{abrial-eventB-96}
Abrial, J.R.:
\newblock Extending {B} without changing it (for developing distributed
  systems).
\newblock In: 1st {B} Conference. (1996)  169--190

\bibitem{Hoare69}
Hoare, C.A.R.:
\newblock An axiomatic basis for computer programming.
\newblock Communications of the ACM \textbf{10}(12) (1969)  576580

\bibitem{SlicingDeModelFormelPourVerif}
Br\"uckner, I., Wehrheim, H.:
\newblock {Slicing an Integrated Formal Method for Verification}.
\newblock In Lau, K.K., Banach, R., eds.: ICFEM'05. Volume 3785 of LNCS.,
  Springer (November 2005)  360--374

\bibitem{bjk00}
Bellegarde, F., Julliand, J., Kouchnarenko, O.:
\newblock Ready-simulation is not ready to express a modular refinement
  relation.
\newblock In: FASE'2000. Volume 1783 of LNCS. (2000)  266--283

\bibitem{jmt-ast08-langage}
Julliand, J., Masson, P.A., Tissot, R.:
\newblock Generating security tests in addition to functional tests.
\newblock In: AST'08, {ACM} Press (2008)  41--44

\bibitem{Thi03}
Thimbleby, H.:
\newblock The directed chinese postman problem.
\newblock Software: Practice and Experience \textbf{33}(11) (2003)  1081--1096

\bibitem{bbjm09}
Bouquet, F., Bu\'e, P.C., Julliand, J., Masson, P.A.:
\newblock G\'en\'eration de tests \`a partir de crit\`eres dynamiques de
  s\'election et par abstraction.
\newblock In: AFADL'09, Toulouse, France (January 2009)  161--176

\bibitem{FriedmanHNS02}
Friedman, G., Hartman, A., Nagin, K., Shiran, T.:
\newblock Projected state machine coverage for software testing.
\newblock In: ISSTA. (2002)  134--143

\bibitem{DickF93}
Dick, J., Faivre, A.:
\newblock Automating the generation and sequencing of test cases from
  model-based specifications.
\newblock In: FME'93. (1993)  268--284

\bibitem{GS97}
Graf, S., Saidi, H.:
\newblock Construction of abstract state graphs with {PVS}.
\newblock In: CAV'97. Volume 1254 of LNCS. (1997)

\bibitem{BLO98}
Bensalem, S., Lakhnech, Y., Owre, S.:
\newblock Computing abstractions of infinite state systems compositionally and
  automatically.
\newblock In: CAV'98. Volume 1427 of LNCS., Springer (1998)

\bibitem{CU98}
Colon, M., Uribe, T.:
\newblock Generating finite-state abstractions of reactive systems using
  decision procedures.
\newblock In: CAV'98. Volume 1427 of LNCS. (1998)

\bibitem{CABN97}
Chan, W., Anderson, R., Beame, P., Notkin, D.:
\newblock {Combining Constraint Solving and Symbolic Model Checking for a Class
  of Systems with Non-Linear Constraints}.
\newblock In: CAV'97. Volume 1254 of LNCS., Springer (1997)

\bibitem{CGL94}
Clarke, E., Grumberg, O., Long, D.:
\newblock {Model Checking and Abstraction}.
\newblock TOPLAS'94, ACM Transactions on Programming Languages and Systems
  \textbf{16}(5) (1994)  1512--1542

\bibitem{SUM99}
Sipma, H., Uribe, T., Manna, Z.:
\newblock Deductive model checking.
\newblock Formal Methods in System Design \textbf{15}(1) (1999)  49--74

\bibitem{NK00}
Namjoshi, K.S., Kurshan, R.P.:
\newblock Syntactic program transformations for automatic abstraction.
\newblock In: CAV'00. Volume 1855 of LNCS., Springer (2000)  435--449

\end{thebibliography}
